\documentclass[twocolumn]{aastex61}

\usepackage{natbib,aas_macros} 
\usepackage{hyperref}
\usepackage{graphicx}	
\usepackage{amsmath,amssymb} 
\usepackage{xspace}
\usepackage{enumerate}
\usepackage{fp} 
\usepackage{lineno}
\usepackage{subfigure}

\newcommand{\ho}{$H_0$\xspace} 
\newcommand{\cchp}{\citetalias{cchp2proposal}\xspace}   
\newcommand{\hst}{\emph{HST}\xspace}
\newcommand{\ic}{IC\,1613\xspace}
\newcommand{\ngctwofour}{NGC\,4424\xspace}
\newcommand{\ngctwosix}{NGC\,4526\xspace}
\newcommand{\ngcthreesix}{NGC\,4536\xspace}
\newcommand{\sne}{SN~Ia\xspace}

\newcommand{\mTRGBngctwofour}{27.028}
\newcommand{\mTRGBngctwosix}{27.012}
\newcommand{\mTRGBngcthreesix}{27.015}

\newcommand{\mTRGBngctwofourSTATerr}{0.0128635}
\newcommand{\mTRGBngctwosixSTATerr}{0.008585}
\newcommand{\mTRGBngcthreesixSTATerr}{0.0147473}

\newcommand{\mTRGBngctwofourSYSerr}{0.00123215}
\newcommand{\mTRGBngctwosixSYSerr}{0.00310326}
\newcommand{\mTRGBngcthreesixSYSerr}{0.0072517}
 
\newcommand{\trgblum}{-4.00}
\newcommand{\trgblumstaterr}{0.03} 
\newcommand{\trgblumsyserr}{0.05}

\newcommand{\Iextinctionngctwofour}{0.032} 
\newcommand{\Iextinctionngctwosix}{0.034} 
\newcommand{\Iextinctionngcthreesix}{0.028} 

\FPeval{\Iextinctionerr}{0.03/2}
\newcommand{\ZPerr}{0.02}
\newcommand{\EEerr}{0.02}
\newcommand{\Apcorrerr}{0.01}

\FPeval{\CalibrationErr}{ (\ZPerr^2+\EEerr^2+\Apcorrerr^2)^0.5 }

\FPeval{\ngctwofourDM}{\mTRGBngctwofour-\Iextinctionngctwofour-\trgblum}
\FPeval{\ngctwosixDM}{\mTRGBngctwosix-\Iextinctionngctwosix-\trgblum}
\FPeval{\ngcthreesixDM}{\mTRGBngcthreesix-\Iextinctionngcthreesix-\trgblum}

\FPeval{\ngctwofourDist}{ 10^( (\ngctwofourDM) /5)/100000 }
\FPeval{\ngctwosixDist}{ 10^( (\ngctwosixDM) /5)/100000 } \FPeval{\ngcthreesixDist}{ 10^( (\ngcthreesixDM) /5)/100000 }

\FPeval{\ngctwofourmTRGBerrCMB}{ (\mTRGBngctwofourSTATerr^2+\mTRGBngctwofourSYSerr^2+\CalibrationErr^2)^0.5 }
\FPeval{\ngctwosixmTRGBerrCMB}{ (\mTRGBngctwosixSTATerr^2+\mTRGBngctwosixSYSerr^2+\CalibrationErr^2)^0.5 }
\FPeval{\ngcthreesixmTRGBerrCMB}{ (\mTRGBngcthreesixSTATerr^2+\mTRGBngcthreesixSYSerr^2+\CalibrationErr^2)^0.5 }

\FPeval{\ngctwofourCMBstaterr}{(\trgblumstaterr^2+\mTRGBngctwofourSTATerr^2)^0.5  }
\FPeval{\ngctwofourCMBsyserr}{(\trgblumsyserr^2+\mTRGBngctwofourSYSerr^2+(\Iextinctionngctwofour/2)^2+\CalibrationErr^2)^0.5  }
\FPeval{\ngctwofourCMBerr}{ (\ngctwofourCMBstaterr^2+\ngctwofourCMBsyserr^2+\Iextinctionerr^2)^0.5 }
\FPeval{\ngctwosixCMBstaterr}{(\trgblumstaterr^2+\mTRGBngctwosixSTATerr^2)^0.5  }
\FPeval{\ngctwosixCMBsyserr}{(\trgblumsyserr^2+\mTRGBngctwosixSYSerr^2+(\Iextinctionngctwosix/2)^2+\CalibrationErr^2)^0.5  }
\FPeval{\ngctwosixCMBerr}{ (\ngctwosixCMBstaterr^2+\ngctwosixCMBsyserr^2+\Iextinctionerr^2)^0.5 }
\FPeval{\ngcthreesixCMBstaterr}{(\trgblumstaterr^2+\mTRGBngcthreesixSTATerr^2)^0.5  }
\FPeval{\ngcthreesixCMBsyserr}{(\trgblumsyserr^2+\mTRGBngcthreesixSYSerr^2+(\Iextinctionngcthreesix/2)^2+\CalibrationErr^2)^0.5  }
\FPeval{\ngcthreesixCMBerr}{ (\ngcthreesixCMBstaterr^2+\ngcthreesixCMBsyserr^2+\Iextinctionerr^2)^0.5 }

\FPeval\ngctwofourDistupperrdiststat{ 10^( (\ngctwofourDM+\ngctwofourCMBstaterr) /5)/100000 }
\FPeval\ngctwofourDistlowerdiststat{ 10^( (\ngctwofourDM-\ngctwofourCMBstaterr) /5)/100000 }
\FPeval\ngctwofourDiststaterr{ 0.5*(\ngctwofourDistupperrdiststat - \ngctwofourDistlowerdiststat) }
\FPeval\ngctwosixDistupperrdiststat{ 10^( (\ngctwosixDM+\ngctwosixCMBstaterr) /5)/100000 }
\FPeval\ngctwosixDistlowerdiststat{ 10^( (\ngctwosixDM-\ngctwosixCMBstaterr) /5)/100000 }
\FPeval\ngctwosixDiststaterr{ 0.5*(\ngctwosixDistupperrdiststat - \ngctwosixDistlowerdiststat) }
\FPeval\ngcthreesixDistupperrdiststat{ 10^( (\ngcthreesixDM+\ngcthreesixCMBstaterr) /5)/100000 }
\FPeval\ngcthreesixDistlowerdiststat{ 10^( (\ngcthreesixDM-\ngcthreesixCMBstaterr) /5)/100000 }
\FPeval\ngcthreesixDiststaterr{ 0.5*(\ngcthreesixDistupperrdiststat - \ngcthreesixDistlowerdiststat) }
\FPeval\ngctwofourDistupperrdistsys{ 10^( (\ngctwofourDM+\ngctwofourCMBsyserr) /5)/100000 }
\FPeval\ngctwofourDistlowerdistsys{ 10^( (\ngctwofourDM-\ngctwofourCMBsyserr) /5)/100000 }
\FPeval\ngctwofourDistsyserr{ 0.5*(\ngctwofourDistupperrdistsys - \ngctwofourDistlowerdistsys) }
\FPeval\ngctwosixDistupperrdistsys{ 10^( (\ngctwosixDM+\ngctwosixCMBsyserr) /5)/100000 }
\FPeval\ngctwosixDistlowerdistsys{ 10^( (\ngctwosixDM-\ngctwosixCMBsyserr) /5)/100000 }
\FPeval\ngctwosixDistsyserr{ 0.5*(\ngctwosixDistupperrdistsys - \ngctwosixDistlowerdistsys) }
\FPeval\ngcthreesixDistupperrdistsys{ 10^( (\ngcthreesixDM+\ngcthreesixCMBsyserr) /5)/100000 }
\FPeval\ngcthreesixDistlowerdistsys{ 10^( (\ngcthreesixDM-\ngcthreesixCMBsyserr) /5)/100000 }
\FPeval\ngcthreesixDistsyserr{ 0.5*(\ngcthreesixDistupperrdistsys - \ngcthreesixDistlowerdistsys) }

\FPeval\ngctwofourDisterr{(\ngctwofourDiststaterr^2+\ngctwofourDistsyserr^2)^0.5}
\FPeval\ngctwofourDisterrPCent{(\ngctwofourDiststaterr^2+\ngctwofourDistsyserr^2)^0.5/\ngctwofourDist*100}
\FPeval\ngctwosixDisterr{(\ngctwosixDiststaterr^2+\ngctwosixDistsyserr^2)^0.5}
\FPeval\ngctwosixDisterrPCent{(\ngctwosixDiststaterr^2+\ngctwosixDistsyserr^2)^0.5/\ngctwosixDist*100}
\FPeval\ngcthreesixDisterr{(\ngcthreesixDiststaterr^2+\ngcthreesixDistsyserr^2)^0.5}
\FPeval\ngcthreesixDisterrPCent{(\ngcthreesixDiststaterr^2+\ngcthreesixDistsyserr^2)^0.5/\ngctwosixDist*100}

\FPeval{\IextinctionngctwofourROUNDED}{round(\Iextinctionngctwofour,2)}
\FPeval{\IextinctionngctwosixROUNDED}{round(\Iextinctionngctwosix,2)}
\FPeval{\IextinctionngcthreesixROUNDED}{round(\Iextinctionngcthreesix,2)}

\FPeval{\mTRGBngctwofourROUNDED}{round(\mTRGBngctwofour,2)}
\FPeval{\mTRGBngctwosixROUNDED}{round(\mTRGBngctwosix,2)}
\FPeval{\mTRGBngcthreesixROUNDED}{round(\mTRGBngcthreesix,2)}
\FPeval{\mTRGBngctwofourSTATerrROUNDED}{round(\mTRGBngctwofourSTATerr,2)}
\FPeval{\mTRGBngctwosixSTATerrROUNDED}{round(\mTRGBngctwosixSTATerr,2)}
\FPeval{\mTRGBngcthreesixSTATerrROUNDED}{round(\mTRGBngcthreesixSTATerr,2)}
\FPeval{\mTRGBngctwofourSYSerrROUNDED}{round(\mTRGBngctwofourSYSerr,2)}
\FPeval{\mTRGBngctwosixSYSerrROUNDED}{round(\mTRGBngctwosixSYSerr,2)}
\FPeval{\mTRGBngcthreesixSYSerrROUNDED}{round(\mTRGBngcthreesixSYSerr,2)}

\FPeval{\ngctwofourmTRGBerrCMBROUNDED}{ round(\ngctwofourmTRGBerrCMB,2) }
\FPeval{\ngctwosixmTRGBerrCMBROUNDED}{ round(\ngctwosixmTRGBerrCMB,2) }
\FPeval{\ngcthreesixmTRGBerrCMBROUNDED}{ round(\ngcthreesixmTRGBerrCMB,2) }

\FPeval{\ngctwofourDMROUNDED}{round(\ngctwofourDM,2)}
\FPeval{\ngctwosixDMROUNDED}{round(\ngctwosixDM,2)}
\FPeval{\ngcthreesixDMROUNDED}{round(\ngcthreesixDM,2)}

\FPeval{\ngctwofourCMBstaterrROUNDED}{ round(\ngctwofourCMBstaterr,2)  }
\FPeval{\ngctwofourCMBsyserrROUNDED}{ round(\ngctwofourCMBsyserr,2) }
\FPeval{\ngctwofourCMBerrROUNDED}{ round(\ngctwofourCMBerr,2) }
\FPeval{\ngctwosixCMBstaterrROUNDED}{ round(\ngctwosixCMBstaterr,2)  }
\FPeval{\ngctwosixCMBsyserrROUNDED}{ round(\ngctwosixCMBsyserr,2) }
\FPeval{\ngctwosixCMBerrROUNDED}{ round(\ngctwosixCMBerr,2) }
\FPeval{\ngcthreesixCMBstaterrROUNDED}{ round(\ngctwofourCMBstaterr,2)  }
\FPeval{\ngcthreesixCMBsyserrROUNDED}{ round(\ngctwofourCMBsyserr,2) }
\FPeval{\ngcthreesixCMBerrROUNDED}{ round(\ngcthreesixCMBerr,2) }

\FPeval{\ngctwofourDistROUNDED}{round(\ngctwofourDist,1)}
\FPeval{\ngctwosixDistROUNDED}{round(\ngctwosixDist,1)}
\FPeval{\ngcthreesixDistROUNDED}{round(\ngcthreesixDist,1)}

\FPeval{\ngctwofourDiststaterrROUNDED}{ round(\ngctwofourDiststaterr,1) }
\FPeval{\ngctwosixDiststaterrROUNDED}{ round(\ngctwosixDiststaterr,1) }
\FPeval{\ngcthreesixDiststaterrROUNDED}{ round(\ngcthreesixDiststaterr,1) }
\FPeval{\ngctwofourDistsyserrROUNDED}{ round(\ngctwofourDistsyserr,1) }
\FPeval{\ngctwosixDistsyserrROUNDED}{ round(\ngctwosixDistsyserr,1) }
\FPeval{\ngcthreesixDistsyserrROUNDED}{ round(\ngcthreesixDistsyserr,1) }

\FPeval\ngctwofourDisterrROUNDED{ round(\ngctwofourDisterr,1) }
\FPeval\ngctwofourDisterrPCentROUNDED{round(\ngctwofourDisterrPCent,1)}
\FPeval\ngctwosixDisterrROUNDED{ round(\ngctwosixDisterr,1) }
\FPeval\ngctwosixDisterrPCentROUNDED{round(\ngctwosixDisterrPCent,1)}
\FPeval\ngcthreesixDisterrROUNDED{ round(\ngcthreesixDisterr,1) }
\FPeval\ngcthreesixDisterrPCentROUNDED{round(\ngcthreesixDisterrPCent,1)}

\shorttitle{TRGB distances to \ngctwofour, \ngctwosix, and \ngcthreesix}
\shortauthors{Hatt et al.}

\defcitealias{cchp2proposal}{CCHP}
\defcitealias{2016ApJ...832..210B}{Paper II}
\defcitealias{hatt17}{Paper II}
\defcitealias{2018ApJ...852...60J}{Paper III}

\begin{document}

\title{\textit{The Carnegie-Chicago Hubble Program.} IV. THE DISTANCES TO\\ \ngctwofour, \ngctwosix, and \ngcthreesix VIA THE TIP OF THE RED GIANT BRANCH\footnote{Based on observations made with the NASA/ESA \emph{Hubble Space Telescope} (\hst), obtained at the Space Telescope Science Institute, which is operated by the Association of Universities for Research in Astronomy, Inc., under NASA contract NAS 5-26555. These observations are associated with program \#13691. Presented as part of a dissertation to the Department of Astronomy and Astrophysics, The University of Chicago, in partial fulfillment of the requirements for the Ph.D. degree.}}


\author[0000-0003-2767-2379]{Dylan~Hatt}\affil{Department of Astronomy \& Astrophysics, University of Chicago, 5640 South Ellis Avenue, Chicago, IL 60637}\email{dhatt@uchicago.edu}

\author[0000-0003-3431-9135]{Wendy~L.~Freedman}\affil{Department of Astronomy \& Astrophysics, University of Chicago, 5640 South Ellis Avenue, Chicago, IL 60637}

\author{Barry~F.~Madore}\affil{Department of Astronomy \& Astrophysics, University of Chicago, 5640 South Ellis Avenue, Chicago, IL 60637}\affil{Observatories of the Carnegie Institution for Science 813 Santa Barbara St., Pasadena, CA~91101}


\author[0000-0002-1691-8217]{Rachael L. Beaton}\altaffiliation{Hubble Fellow}\altaffiliation{Carnegie-Princeton Fellow}\affiliation{Department of Astrophysical Sciences, Princeton University, 4 Ivy Lane, Princeton, NJ~08544}

\author{Taylor~J.~Hoyt}\affil{Department of Astronomy \& Astrophysics, University of Chicago, 5640 South Ellis Avenue, Chicago, IL 60637}

\author[0000-0002-2502-0070]{In~Sung~Jang}\affil{Leibniz-Institut f\"{u}r Astrophysik Potsdam, D-14482 Potsdam, Germany}

\author{ Myung~Gyoon~Lee}\affil{Department of Physics \& Astronomy, Seoul National University, Gwanak-gu, Seoul 151-742, Korea}

\author{Andrew~J.~Monson}\affil{Department of Astronomy \& Astrophysics, Pennsylvania State University, 525 Davey Lab, University Park, PA 16802}

\author{Jeffrey~A.~Rich}\affil{Observatories of the Carnegie Institution for Science 813 Santa Barbara St., Pasadena, CA~91101}

\author{Victoria~Scowcroft}\affil{Department of Physics, University of Bath, Claverton Down, Bath, BA2 7AY, United Kingdom}

\author{Mark~Seibert}\affil{Observatories of the Carnegie Institution for Science 813 Santa Barbara St., Pasadena, CA~91101}


\begin{abstract}

The \emph{Carnegie-Chicago Hubble Program} (\cchp) is undertaking a re-calibration of the extragalactic distance scale, using Type Ia supernovae that are tied to Tip of the Red Giant Branch (TRGB) distances to local galaxies. 
We present here deep \emph{Hubble Space Telescope} (\hst) ACS/WFC imaging of the resolved stellar populations in the metal-poor halos of the \sne host galaxies \ngctwofour, \ngctwosix, and \ngcthreesix. These three Virgo constellation galaxies are prime targets for calibrating the extragalactic distance scale given their relative proximity in the local Universe and their low line-of-sight reddenings. Anchoring the TRGB zero-point to the geometric distance to the Large Magellanic Cloud via detached eclipsing binaries,
we measure extinction-corrected distance moduli of $\ngctwofourDMROUNDED\pm\ngctwofourCMBstaterrROUNDED_{stat}\pm\ngctwofourCMBsyserrROUNDED_{sys}$~mag, $\ngctwosixDMROUNDED\pm\ngctwosixCMBstaterrROUNDED_{stat}\pm\ngctwosixCMBsyserrROUNDED_{sys}$~mag, and $\ngcthreesixDMROUNDED\pm\ngcthreesixCMBstaterrROUNDED_{stat}\pm\ngcthreesixCMBsyserrROUNDED_{sys}$~mag for \ngctwofour, \ngctwosix, and \ngcthreesix, respectively, or $\ngctwofourDistROUNDED\pm\ngctwofourDiststaterrROUNDED_{stat}\pm\ngctwofourDistsyserrROUNDED_{sys}$~Mpc, $\ngctwosixDistROUNDED\pm\ngctwosixDiststaterrROUNDED_{stat}\pm\ngctwosixDistsyserrROUNDED_{sys}$~Mpc, and $\ngcthreesixDistROUNDED\pm\ngcthreesixDiststaterrROUNDED_{stat}\pm\ngcthreesixDistsyserrROUNDED_{sys}$~Mpc. For these three galaxies, the distances are the first based on the TRGB, and for \ngctwofour and \ngctwosix, they are the  highest precision distances published to date, each measured to 3\%.
Finally, we report good agreement between our TRGB distances and the available Cepheid distances for \ngctwofour and \ngcthreesix, demonstrating consistency between the distance scales currently derived from stars of Population I and II.

\end{abstract}



\section{Introduction} \label{sec:intro}
At present there remains a significant tension in the value of \ho as determined by direct methods  \citep[e.g.,][]{2012ApJ...758...24F,2016ApJ...826...56R} and by indirect methods, such as modeling of the Cosmic Microwave Background \citep[e.g.,][]{2011ApJS..192...18K,planck16}. The need for improved systematic accuracy in the direct distance ladder has motivated the \emph{Carnegie-Chicago Hubble Program} (\cchp), which is calibrating anew the \sne extragalactic distance scale, using Population (Pop) II stars. This route is independent of and parallel to the traditional Pop~I, Cepheid-based distance scale, and can, therefore,  provide insight into the current divide in the measurement of \ho.

Pop~II stars have numerous advantages over their Pop~I counterparts as distance indicators. In the first instance, Pop~II stars are found  in the gas- and dust-free, outer halos of galaxies, where source crowding and confusion is also less than that of the disk and where the uncertainty associated with line-of-sight reddening is naturally minimized. 

The primary focus of the \cchp is the Tip of the Red Giant Branch (TRGB) Method, which has been well-studied both theoretically \citep[][]{ 1983ARA&A..21..271I,1992ApJ...400..280R} and empirically \citep[an overview of recent calibration efforts is given in][]{bel08}. In brief, the He-flash, marks the point of rapid departure of stars from the high-luminosity (hydrogen shell-burning) TRGB over to bluer colors (higher temperatures, smaller radii) and onto the lower-luminosity (helium core-burning) Horizontal Branch. This phase is initiated by the lifting of degeneracy in the helium cores of red giant branch (RGB) stars having total masses of $\lesssim1.8~M_{\sun}$ \citep{salaris_stellar_pop}. The peak luminosity in this phase is insensitive to stellar metallicities as high as $[\mathrm{Fe/H}]=-0.3$ \citep{2004ApJ...606..869B}.  

This \cchp paper is the second in detailing the measurement of TRGB distances to nine nearby galaxies containing a cumulative 12 \sne. Previously we have published a distance to NGC~1365 \citep[][Paper III]{2018ApJ...852...60J}, host to the \sne SN~2012fr.
In addition, we have measured the distance to the Local Group galaxy \ic \citep[][hereafter, Paper II]{hatt17}. \ic and NGC~1365 represent the extremes in distance for galaxies studied in the \cchp, approximately 730~kpc and 18.1~Mpc, respectively. A summary of science goals and expected error budgets, including calibration of the RR Lyrae distance scale, are given in \citet[Paper I]{2016ApJ...832..210B}. 

In this paper we present TRGB distances to three \sne host galaxies---\ngctwofour, \ngctwosix, and \ngcthreesix---using deep \emph{Hubble Space Telescope} (\hst) imaging of their halos. \ngctwofour is a barred spiral galaxy that was host to the recent \sne, SN~2012cg \citep{2012ATel.4115....1C}. \ngctwosix is an S0/lenticular galaxy in the Virgo Cluster and was host to SN~1994D \citep{1994IAUC.5946....2T}. Finally, \ngcthreesix is a spiral galaxy that was host to SN~1981B \citep{1983ApJ...270..123B}. Earlier work suggests that these galaxies all lie at a similar distance of $\sim15$~Mpc, and they can therefore serve as important calibrators for the \sne distance scale in the local Universe. 

Moreover, the results presented here are of particular importance to the \sne distance scale because (1) they are the first TRGB distances for these galaxies and (2), in the case of \ngctwofour and \ngctwosix, these are the first distance estimates achieving high-precision (3\%) errors in distance. An increased level of precision for these distances is made possible, in part, by the recent availability of an improved (2.2\% precision) geometric distance to the Large Magellanic Cloud (LMC) derived from detached eclipsing binaries \citep{2013Natur.495...76P}. In the near-term, we anticipate that it will be possible to even more precisely anchor the TRGB distance scale (to 1\% precision) directly within the Milky Way, using \emph{Gaia} parallaxes of RGB stars. At that time, we anticipate that the precision of the results presented here will improve further.

The paper is organized as follows: 
Section \ref{sec:data} describes the observations and photometry;  Section \ref{sec:trgb} presents the analysis of the TRGB, including the estimation of measurement uncertainties and the determination of distances;  Section \ref{sec:dist_compare} places the distances measured here in context with previously-published estimates, especially recent Cepheid-based determinations; and Section \ref{sec:conc} provides a summary and the immediate impact of the results presented in this study.

\section{Data} \label{sec:data}

The image processing and extraction of photometry follow the procedures described in \citetalias{hatt17} and employed for the \sne host galaxy NGC\,1365 in \citetalias{2018ApJ...852...60J}. 
In the following sections, we briefly summarize the observations as well as their reduction and calibration.

\begin{figure*}
\centering
\hspace{-100px}
\mbox{\includegraphics[angle=0,width=0.475\textwidth]{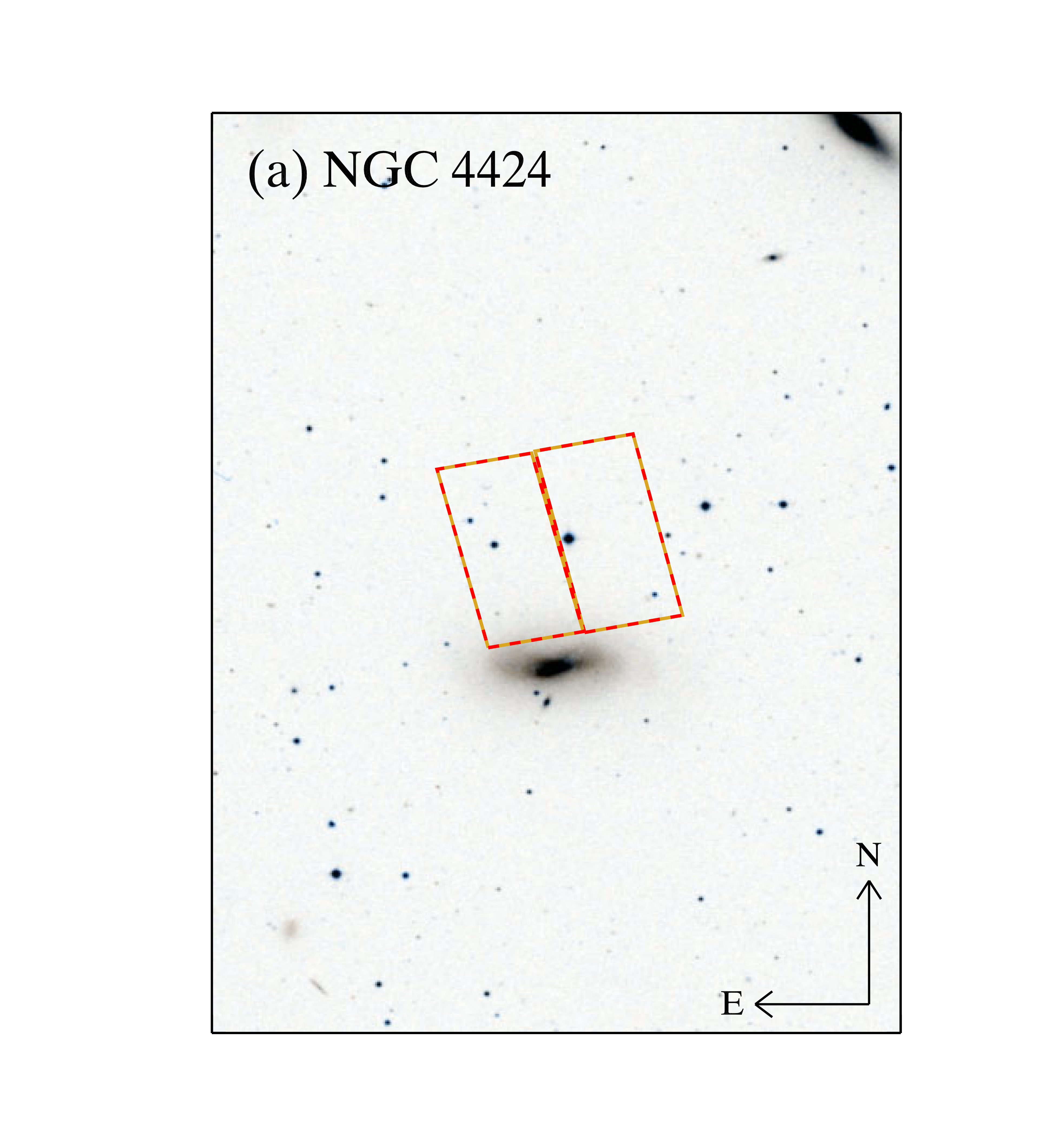}}
\hspace{-75px}
\mbox{\includegraphics[angle=0,width=0.475\textwidth]{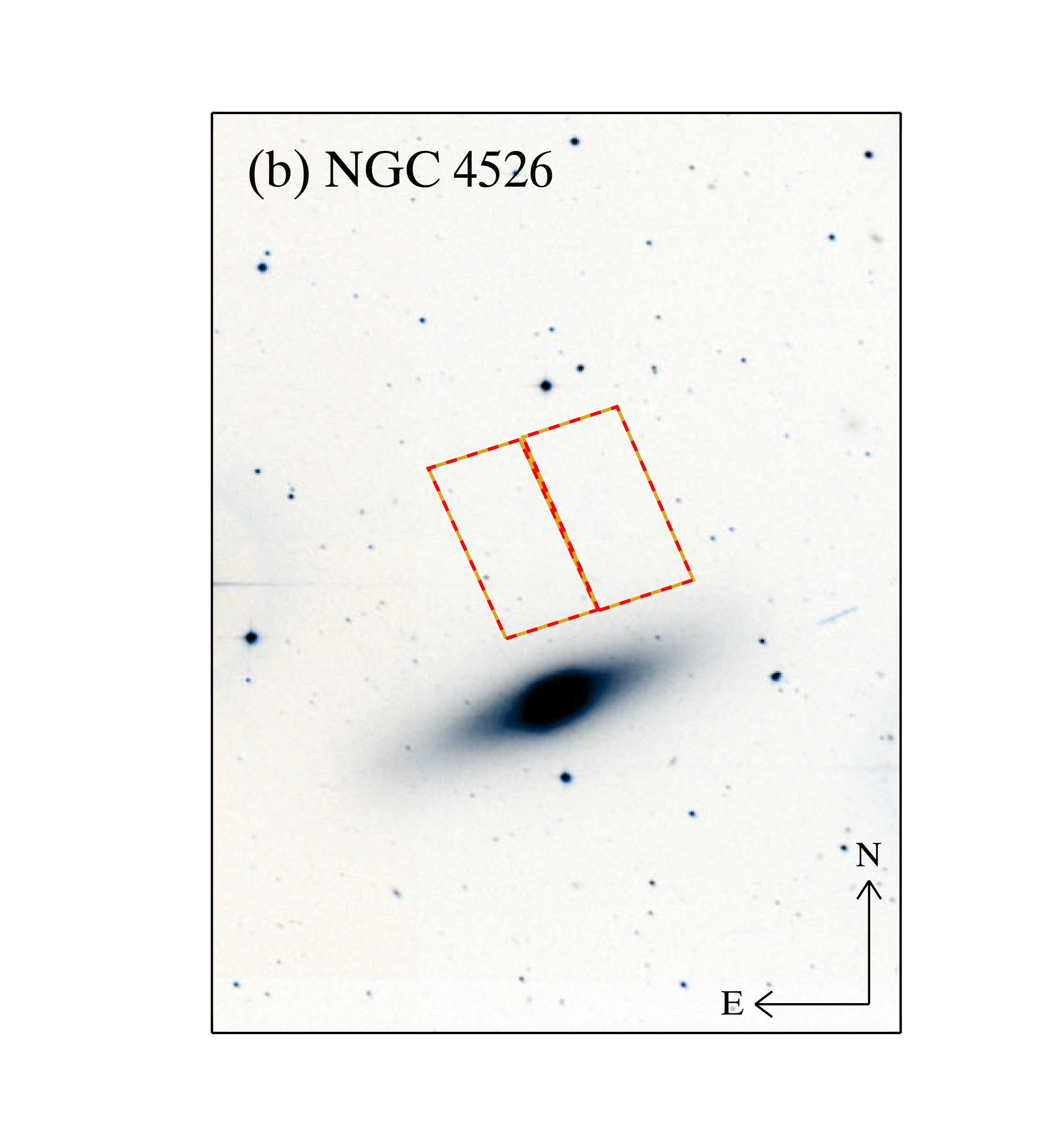}}
\hspace{-75px}
\mbox{\includegraphics[angle=0,width=0.475\textwidth]{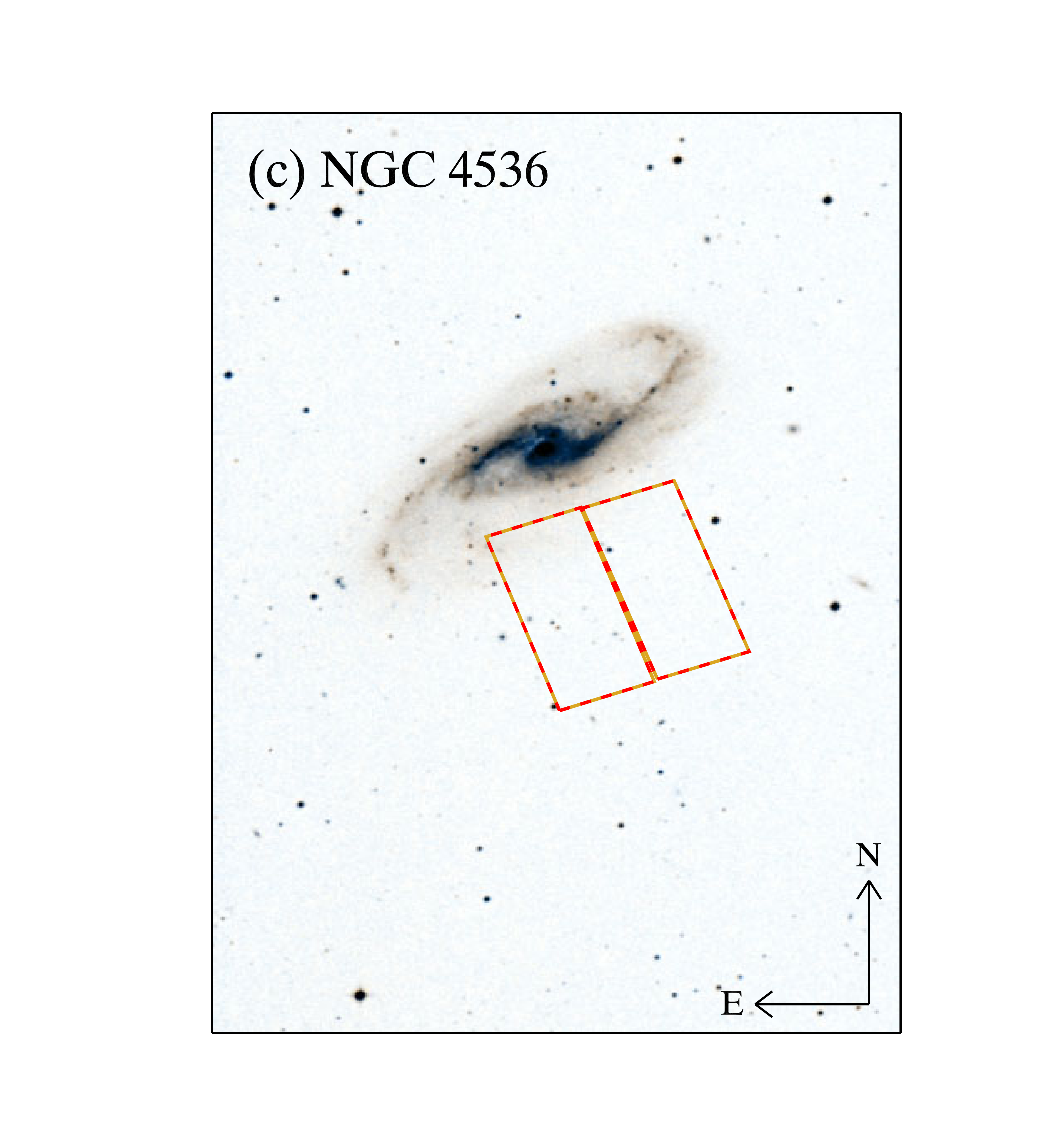}}
\hspace{-85px}
\vspace{-20pt} 
\caption{Locations of \hst ACS/WFC F606W and F814W imaging (boxes) overlaid on an inverted color DSS image for (a) \ngctwofour, (b) \ngctwosix, and (c) \ngcthreesix. Each display spans $12\farcm0\times16\farcm7$ relative to the $3\farcm37\times3\farcm37$ ACS/WFC imaging.
\label{fig:imaging}}
\end{figure*}

\begin{deluxetable*}{cccccccc} 
\tabletypesize{\normalsize} 
\tablewidth{0pt} 
\tablecaption{ACS/WFC Observation Summary \label{tbl:obs_sum}} 
\tablehead{ 
\colhead{Target} &
\colhead{Dates} &
\colhead{Filters [no. obs.~$\times$~exposure time (s)]} &
\colhead{$\alpha~(2000)$} &
\colhead{$\delta~(2000)$} &
\colhead{Field Size} 
}
\startdata 
\ngctwofour & 2015 May 18 & F606W $[3\times 1200]$, F814W [$9\times 1200$] &  $12^h 27^m 12^s$ & $+09^\circ 27\arcmin 32\arcsec$ & $3\farcm37\times 3\farcm37\xspace$  \\
\ngctwosix & 2015 Jun 13 & F606W [$3\times 1200$], F814W [$9\times 1200$] &  $12^h 34^m 04^s$ & $+07^\circ 45\arcmin 11\arcsec$ & $3\farcm37\times 3\farcm37\xspace$ \\
\ngcthreesix & 2015 Dec 14-15 & F606W [$3\times 1200$], F814W [$9\times 1200$] &  $12^h 34^m 21^s$ & $+02^\circ 08\arcmin 41\arcsec$ & $3\farcm37\times 3\farcm37\xspace$  \\
\enddata 
\tablecomments{See also \autoref{fig:imaging} for imaging coverage.} 
\end{deluxetable*} 

\subsection{Observations}

Imaging in this study made use of the \hst Advanced Camera for Surveys with the Wide-Field Channel (ACS/WFC). Observations were specifically designed to target galaxy halos, where the selection of the fields is described in \citetalias{2016ApJ...832..210B}. In summary, these fields were chosen: i) to avoid disks (if any), young (blue) populations and/or tidal structures; ii) to straddle the WISE W1 25-26~mag isophote and the GALEX NUV 27-28~mag isophote, if applicable; and iii) to be aligned along the minor axes so as to maximize the number of uncrowded halo stars within each field-of-view. \autoref{fig:imaging} displays the imaging coverage for this study, and Table \ref{tbl:obs_sum} provides a summary of the observations. 

For each galaxy, 6 orbits were devoted to 3 F606W and 9 F814W exposures, each of which were approximately $1200$~sec in length. Exposure time calculations were optimized to achieve an approximate S/N of 10, in F814W, at the anticipated magnitude of the TRGB. The image products used here were FLC data files, obtained from the {\it Space Telescope Science Institute} website. As received, these images are calibrated, flat-fielded, and CTE-corrected. Each frame was then multiplied by its corresponding Pixel Area Map\footnote{\url{http://www.stsci.edu/hst/acs/analysis/PAMS}} to account for flux differences due to ACS/WFC geometric distortions.

\subsection{Photometry}

We derived instrumental magnitudes from individual images through point-spread-function (PSF) fitting using the \textsc{DAOPHOT} software \citep{1987PASP...99..191S}.
The PSFs for F606W and F814W were created using synthetic star grids generated using the \hst Tiny Tim PSF modeling software \citep{2011SPIE.8127E..0JK}. 
A direct test of the Tiny Tim PSFs against direct frame-by-frame PSF modeling with isolated, bright stars is described in \citetalias{hatt17} and was found to agree to within the quoted photometric uncertainties. 
Images were aligned from preliminary catalogs using \textsc{DAOMATCH}/ \textsc{DAOMASTER} \citep{1987PASP...99..191S} then simultaneously photometered using  the \textsc{ALLFRAME} software \citep{1994PASP..106..250S} with a `master source list' from a co-add of all images.

\subsection{Calibration of photometry} \label{sssec:hstcal}

\begin{deluxetable*}{ccccccccc} 
\tabletypesize{\normalsize} 
\tablewidth{0pt} 
\tablecaption{Measured aperture corrections at 0\farcs5\label{tbl:phot_cal}} 
\tablehead{ 
\colhead{Target} &
\multicolumn{2}{c}{chip 1}  &
\multicolumn{2}{c}{chip 2}   \\
\colhead{} &
\colhead{F606W (No.)} & \colhead{F814W (No.)} &
\colhead{F606W (No.)} & \colhead{F814W (No.)}
}
\startdata 
\ngctwofour & $-0.12(26)$ & $-0.13(121)$ & $-0.12(24)$ & $-0.12 (100)$\\
\ngctwosix & $-0.12 (33)$ & $-0.10 (111)$ & $-0.12 (37)$ & $-0.12 (99)$ \\
\ngcthreesix & $-0.08 (29)$ & $-0.06 (90)$ & $-0.10 (45)$ & $-0.07 (142)$ \\
\enddata 
\tablecomments{Number of bright, isolated stars used in the average is given in parentheses.} 
\end{deluxetable*}

Instrumental magnitudes were brought onto the \hst Vega magnitude system using Equations 2 and 4 of \cite{2005PASP..117.1049S}.
The photometric zero-points are 26.405 mag for F606W and 25.517 mag for F814W, accessed on 2018-05-12\footnote{\url https://acszeropoints.stsci.edu/}. We note that there is a systematic uncertainty in the observed flux of Vega, which forms the basis for the ACS zero-point calculation.
As per \cite{2005PASP..117.1049S}, we adopt a conservative 2\% systematic uncertainty in flux (0.02~mag) for these zero-points.

The correction from the PSF magnitudes to the 0\farcs5 aperture magnitudes was calculated for each filter and CCD combination. Due to the small number of bright and isolated stars that are free of cosmic rays (typically $\sim3$ per frame), we have previously adopted a mean value from all frames for each filter and CCD. This approach was shown in \citetalias{hatt17} to be indistinguishable from tailoring corrections to a frame-by-frame basis when bright and isolated stars are abundant. In this work, we have found that the frame-by-frame corrections, obtained through an automated selection of bright and isolated stars, yield the same science result as when using the average correction. In the following, all results use the frame-by-frame corrections. In Table \ref{tbl:phot_cal} we list the averages for brevity, where the standard deviation of values across all frames for a given CCD and filter is typically 0.03~mag. We have found during the study of NGC\,1365 in \citetalias{2018ApJ...852...60J} that independent efforts in the selection of bright, isolated stars resulted in differences at the $\sim0.01$~mag level. In the following, we therefore adopt a $0.01$~mag systematic uncertainty in the photometric calibration due to the aperture correction at $0\farcs5$. We plan to investigate in more detail the uncertainty of this correction in future works.

The 0\farcs5 to infinite aperture correction values are 0.095 and 0.098~mag for the F606W and F814W filters, respectively \citep{boh16}, which were computed from the provided Encircled-Energy (EE) tables. \cite{boh16} quotes a 4\% uncertainty in EE for cool, late-type stars (i.e. RGB stars) that are used to compute these 0\farcs5 to infinite aperture corrections. 
Table 6 of \cite{2005PASP..117.1049S}, however, shows that the total variation in EE due to changes in effective wavelength via spectral type is $<0.01$~mag for WFC at 0\farcs5, implying that the EE for these cooler stars are consistent with hotter stars that are constrained at the $1\%$ level.
In this study we adopt half of the \cite{boh16} estimate, or 2\% error in flux (0.02~mag) as another systematic uncertainty due to the scatter in the measured EE for the RGB stars that are the focus of this work. In total, we have  adopted a combined 0.03~mag systematic uncertainty attributed to the Vegamag zero-points, measured aperture corrections, and the Encircled-Energy of late-type stars.

\subsection{Color-magnitude diagrams}\label{sec:cmds}

\begin{figure*}
\centering
\hspace{-15px}
\mbox{\includegraphics[angle=0,width=0.35\textwidth]{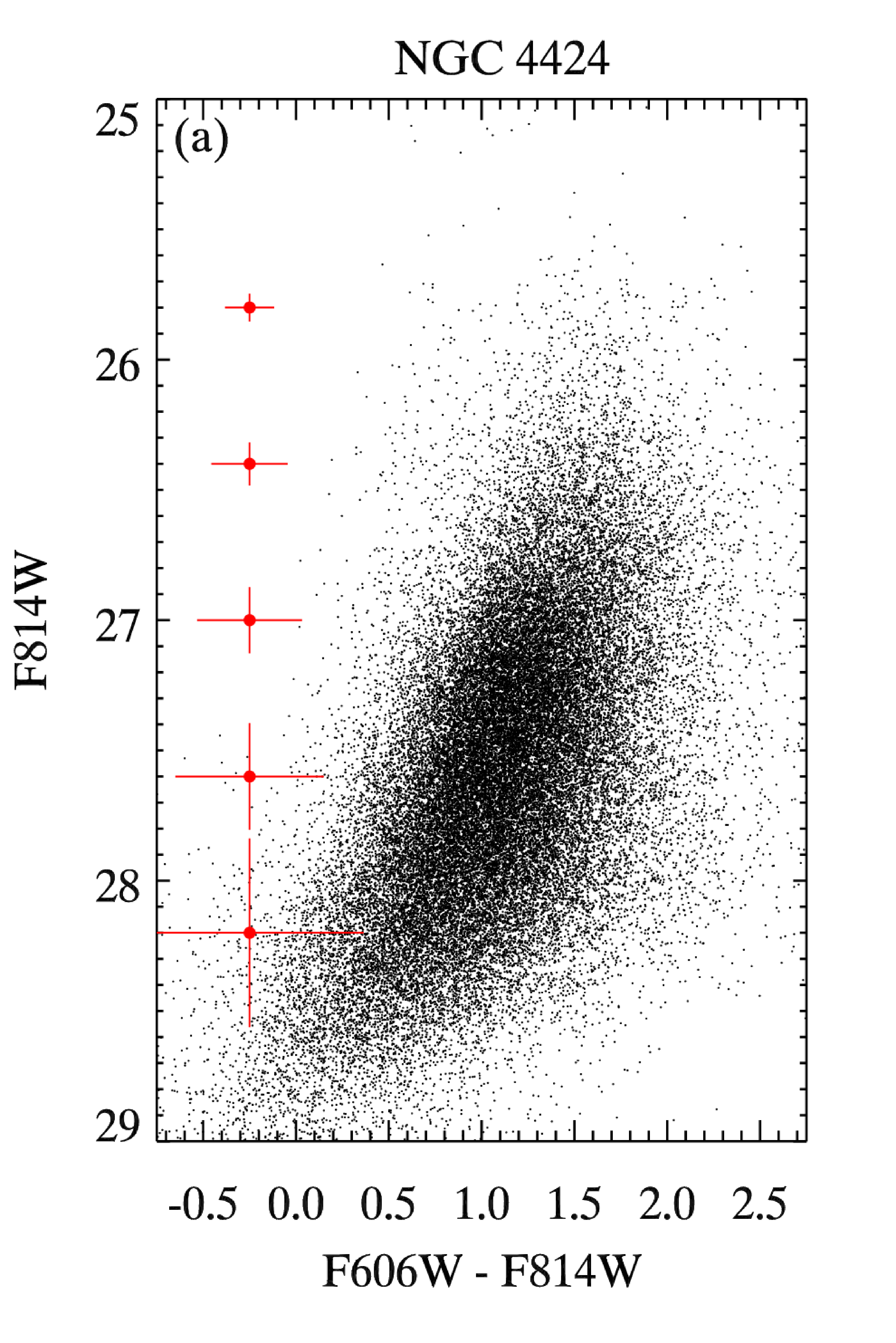}}
\hspace{-15px}
\mbox{\includegraphics[angle=0,width=0.35\textwidth]{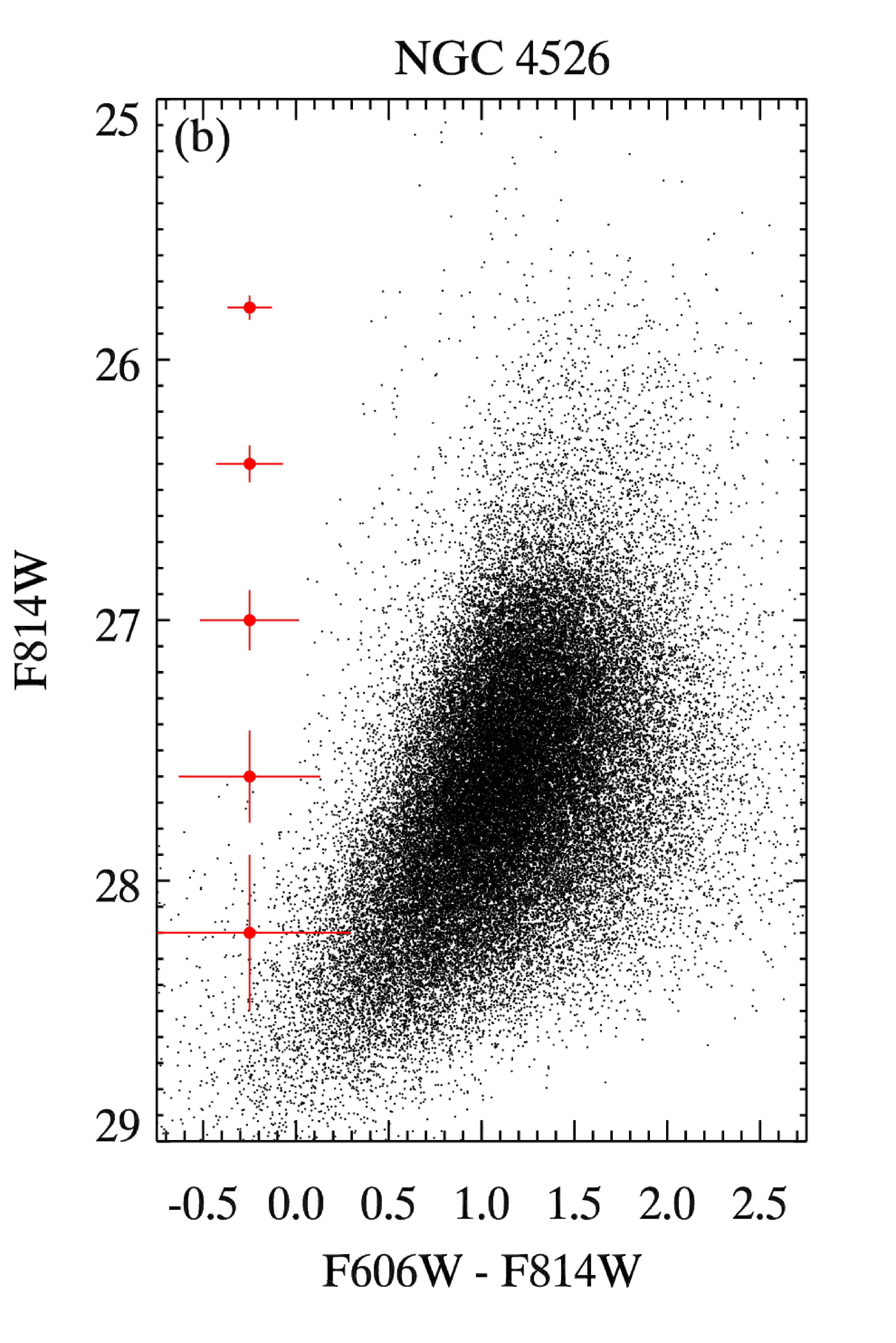}}
\hspace{-15px}
\mbox{\includegraphics[angle=0,width=0.35\textwidth]{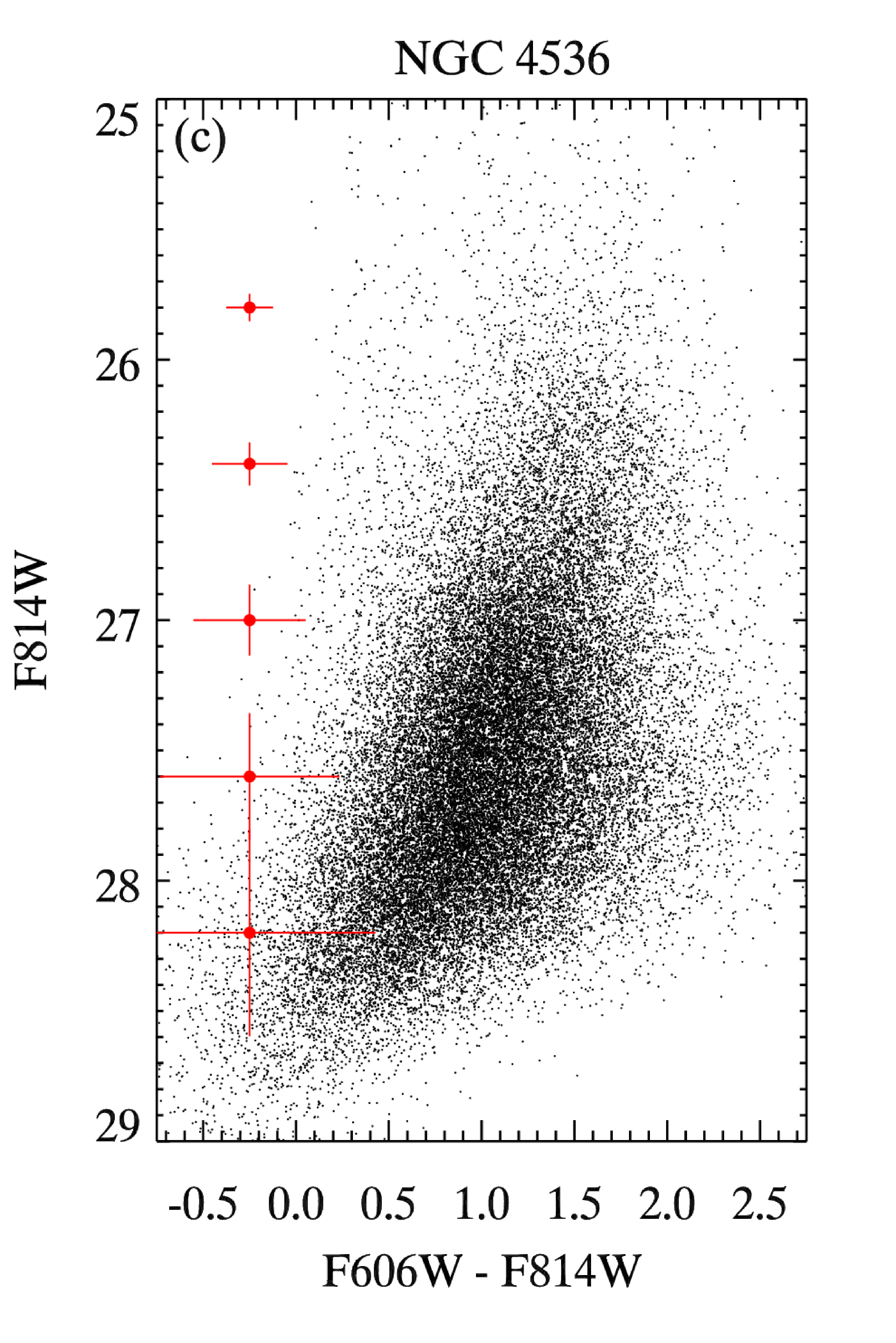}}
\hspace{-15px}
\vspace{-10pt}
\caption{Color-magnitude diagrams of the galaxies \ngctwofour, \ngctwosix, and \ngcthreesix. Likely TP-AGB stars are mixed with Red Giant Branch stars, though a steep jump in the luminosity function of each CMD still marks the location of the TRGB as seen later in this work. Median uncertainties in magnitude and color are shown alongside each CMD. 
\label{fig:cmds}}
\end{figure*}

The calibrated photometry is presented in the form of Color-Magnitude Diagrams (CMDs) in \autoref{fig:cmds}. In each plot, median uncertainties in measured F814W and $\mathrm{F606W}-\mathrm{F814W}$ are shown for a variety of magnitudes. Each CMD shows a `pure halo' component, i.e. only an RGB, mixed with likely thermally-pulsating Asymptotic Giant Branch (TP-AGB) stars above the RGB and early-type AGB (E-AGB) in the same magnitude range as the RGB. In the following, we often refer to the two AGB classes as a single `AGB' component. The small color range for the RGB in each CMD reflects the metal-poor nature of the halos surrounding these galaxies. In a later section, we quantify the effect of contamination by AGB sources as well as assess the level of photometric completeness and crowding/blending through extensive artificial star tests and population models. 
 
\autoref{fig:cmds}a displays the CMD for \ngctwofour. This field has the second-most notable AGB/blended RGB component of the three galaxies in this study, although the TRGB is still readily detected as demonstrated later in this work. The contamination by AGB/blends, or blurring of the TRGB, is partly attributable to crowding and source confusion given the relative proximity of the observations to the galaxy center.

\autoref{fig:cmds}b is the CMD of \ngctwosix, which has the most-populated RGB, and consequently, the most-populated TRGB. It is apparent that there is less of an AGB/blended RGB component compared to \ngctwofour. Given the favorable number of stars defining the tip, the major contribution to the uncertainty in measuring the TRGB is then expected to be photometric errors.

\autoref{fig:cmds}c shows the CMD of \ngcthreesix, which has the most prominent AGB/blended RGB component of the three galaxies studied here. Although the TRGB of \ngcthreesix is more visually obscured in comparison to the previous two cases, the jump in the star counts in the luminosity function at the TRGB is nonetheless readily detectable, as demonstrated in the following Section.

As a special note for \ngcthreesix, we identified a luminous stellar component that spans part the highest star density region of the imaging for chip 2. The feature appears visually consistent with a spiral arm or other stellar feature. This possible spiral arm suggests that the CMD for \ngcthreesix is not a `pure' halo. The impact of the blue sources is quantified later in Section \ref{ssec:dist}.

\section{The Tip of the Red Giant Branch} \label{sec:trgb}

\subsection{Background}

The TRGB is a discontinuity in the RGB luminosity function (LF) for low-mass stars as they rapidly drop in luminosity while evolving over and down onto the Horizontal Branch. This process is triggered by the Helium-flash at a critical core mass, where an electron degenerate equation-of-state simultaneously fixes the star's bolometric luminosity. In the $I$-band (or in the \hst equivalent, F814W filter), the TRGB is remarkably fixed in brightness for observations of metal-poor stars, such as those presented here \citep[although corrections for the higher-metallicity extension have recently been calibrated empirically for ACS/WFC filters by][]{2017ApJ...835...28J}. A detailed physical description of the TRGB and its measurement is given in \citetalias{hatt17}, \citetalias{2018ApJ...852...60J}, and the references therein. In the following, we revisit only major points.

In making our measurements of the brightness of the TRGB, we have used bins of 0.01~mag in width to measure the relative star abundance in the F814W luminosity function (LF). This finely binned LF has then been smoothed using GLOESS (Gaussian-windowed, Locally-Weighted Scatterplot Smoothing), which is an interpolating technique first used in an astrophysical context for Cepheid light curves by \citet{per04}. A further example of GLOESS can also be found in \citet{2017AJ....153...96M} in the discussion of RR Lyrae light curves.

The underlying principle behind GLOESS smoothing is the sequential moving and fitting of a second-order polynomial to the entire data set of points weighted by their Gaussian distance to a given reference point. This polynomial is swept across the interval of interest at arbitrarily small, but user-controlled, steps. The degree of smoothing is controlled by the width of the Gaussian window, which we define as $\sigma_{s}$. On the smoothed LF, we then apply a Sobel filter $[-1, 0, +1]$ \citep[as first introduced for this application by][]{1993ApJ...417..553L}, which is the finite-difference version of the first-derivative for a discrete function. This Sobel kernel is an effective `edge detector' given that it produces the maximum response when the local change in the LF is greatest, i.e. at discontinuities, such as that encountered at the TRGB.

\citetalias{hatt17} describes a procedure using artificial star tests to empirically derive the value for $\sigma_{s}$ that minimizes the the associated statistical (random) and systematic uncertainties in the measurement of the TRGB, which are expected to be unique for a given dataset. In the following section, we create similar artificial star luminosity functions that are tailored to the observations in this study.

\subsection{Artificial star luminosity functions}\label{ssec:artstar_lum}

\begin{figure*}
\centering
\advance\leftskip-1.2cm
\includegraphics[angle=0,width=1.15\textwidth]{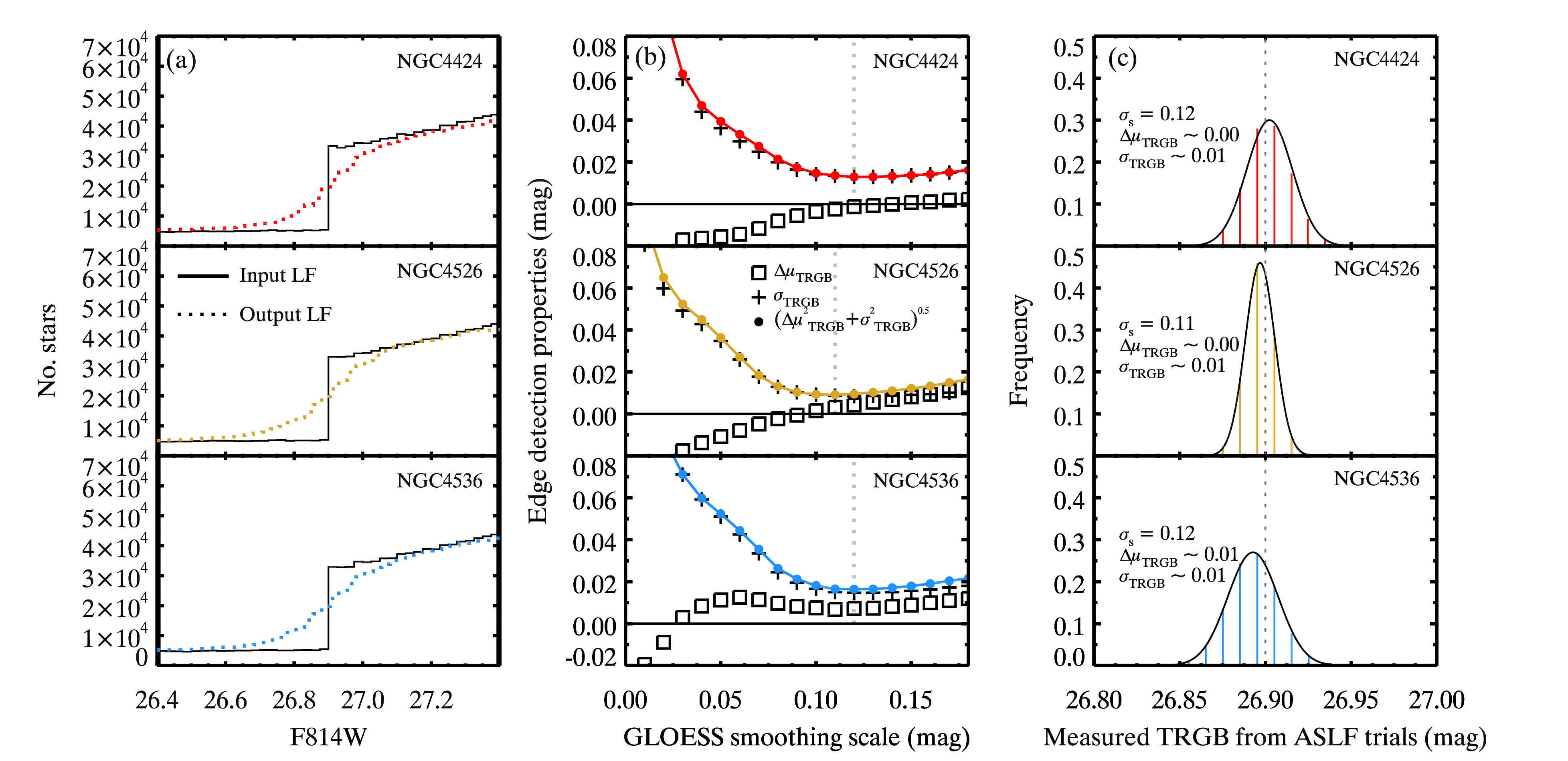}
\caption{Estimating edge detection uncertainties through artificial star tests. Left Panel (a) input (solid) and output (dashed) artificial star AGB+RGB luminosity functions. Each input TRGB is assigned a value $\mathrm{F814W}=26.9$ for the simulations. 
Middle Panel (b) Statistical/random (plus signs), systematic (squares), and combined measurement uncertainties (points and lines) associated with the $[-1,0,+1]$ Sobel edge-detection kernel as a function of GLOESS smoothing scale for each galaxy. A minimum for each galaxy, marked by a dotted gray vertical line, occurs where the smoothing scale is both large enough to reduce Poisson noise in the luminosity functions and still small enough to avoid displacing the true location of the TRGB. The computed optimal level of smoothing is comparable for each galaxy.
Right Panel (c) The distribution of measured TRGB values at the optimal level of GLOESS smoothing. Measured statistical and systematic uncertainties are rounded to the nearest one-hundredth for the plot, which are $< 0.005$~mag is some cases.
\label{fig:artstars}}
\end{figure*}

In order to effectively model the RGB and AGB populations as seen in our observations, we first created an artificial star luminosity function for each galaxy in order to better understand the natural blurring of the TRGB discontinuity due to factors including photometric errors, crowding, and contamination by AGB stars.

As described in \citetalias{hatt17}, we adopted the RGB and AGB luminosity function slopes of $0.3$ dex~mag$^{-1}$ and $0.1$~dex~mag$^{-1}$, respectively.
For each galaxy, $2,000$ stars were placed into each of our ACS/WFC CCD frames at pixel coordinates chosen by randomly sampling from a uniform distribution in $X$ and $Y$ in order to avoid over-crowding in any given realization. Stellar magnitudes were drawn from the RGB and AGB LFs as described above. The RGB LF was chosen to begin at $\mathrm{F814W}=26.9$~mag, and was then sampled 1~mag fainter. The AGB LF was started 1~mag brighter than the input TRGB and extended 1~mag downward through the TRGB, down to faint end of the RGB LF. The color for each star was drawn from a uniform distribution $1.0\leq (\mathrm{F606W}-\mathrm{F814W})\leq1.5$~mag to adequately model the natural breadth of the TRGB due to metallicity, though still within the anticipated metal-poor range where `tip rectification' or metallicity correction tools are unnecessary.  Finally, we set the relative number of RGB to AGB stars below and above the tip to have a population ratio of 4:1. This ratio is a lower limit on the RGB-AGB contrast based on recently published statistics for these populations in local galaxies \citep[e.g.][]{rose_2014}. The simulation was repeated 500 times for each CCD, producing $500\times 2\times 2000=2,000,000$ artificial stars per field/galaxy (see \autoref{fig:artstars}a). 

\subsection{Photometric completeness and crowding}\label{sec:compl}

Based on the photometry of artificial stars, we find that the fraction of recovered stars is primarily dependent on the S/N of the input star rather than the physical location the star in the imaging; moreover, the completeness of stars across our sample of galaxies is fairly stable in the simulated magnitude range. The recovery rate of stars at the faint limit of the artificial star simulations is 96\%, 98\%, and 91\% for \ngctwofour, \ngctwosix, and \ngcthreesix, respectively. In the TRGB magnitude range, the recovery of artificial stars stands at the 98-99\% level. We conclude that source crowding is not a significant factor in recovering stars within our datasets.

Regarding blends, we have performed an estimation of possible blends through surface brightness arguments via \cite{1998AJ....115.2459R} as follows: we divide our imaging into sections, aligned along the minor axes of each galaxy so that we take into account that the surface brightness of the imaging trends with distance from the central galaxy. We compute the total counts in F606W \citep[the $V$-band is used by ][]{1998AJ....115.2459R}, correct to $V$-band magnitudes using Table 22 of \cite{2005PASP..117.1049S},
then transform them into absolute magnitudes using the approximate distance moduli that we measure via the TRGB (distances are derived later in Section \ref{ssec:dist}).

Using units of solar luminosities per pixel, we scale the population counts in the \cite{1998AJ....115.2459R} Table 1 from their $10^5~L_\odot$. The probability of a blend (when the predicted numbers are less than unity) is simply the product of the predicted abundances. We compute the probability of a blend for each region in each galaxy by considering all likely pair combinations of RGBT, E-AGB, and TP-AGB stars \citep[RGBT is defined as an RGB stars within 1 magnitude of the TRGB in][]{1998AJ....115.2459R}. We assume that the photometry in the region at hand is likely compromised, making a measurement of the TRGB difficult, if the probability of a blend exceeds 0.05 (5\%). We find that the cumulative probability exceeds our defined blend threshold in the $10\%$ of the footprint that is closest to the disk of \emph{only} \ngctwofour, especially in the case of chip 2. In the following analysis, we exclude this region.

\subsection{Optimizing the TRGB edge detection}\label{ssec:opt_trgb}

With artificial star luminosity functions in hand that probe the properties of the images used in this study, we now investigate how a randomly generated LF subset---comparable in number counts and slope to the observed data---affects the measurement of the TRGB as a function of the adopted GLOESS smoothing scale.

In order to ensure that our artificial star LFs are reflective of the observed LFs, the relative number of RGB to AGB stars was adjusted so that the slope of the artificial LF at the TRGB mirrored that of the observed LF. This alteration is permissible because our method relies only on the total shape of the LF, i.e. it does not differentiate between the shapes of the individual stellar components. Consequently, the exact slopes or number counts of  AGB and RGB components is unimportant.

We first estimate the number of stars that contribute to the TRGB measurement. We count 2852 (1147), 2054 (2133), and 1121 (3285) stars in chip 1 (chip 2) within a $\pm0.25$~mag interval centered on the estimated TRGB magnitude for \ngctwofour, \ngctwosix, and \ngcthreesix, respectively. We then model the true locations of sources by their Cumulative Distribution Functions (CDFs). Although we do not expect crowding to play a significant role in our TRGB measurement as discussed above, we nonetheless model the locations of sources as closely as possible to embody all possible known and unknown properties of the images. 
We then apply inverse transform sampling of these CDFs, locating the artificial stars that lie closest in $X$ and $Y$, until the number of artificial stars for each CCD matches the observed count. For a fixed GLOESS $\sigma_s$, we run our Sobel edge detector and record the location of the greatest change in the LF. This process was repeated 10,000 times (enough for the TRGB simulations to converge across all 3 galaxies) for smoothing scales of $0.01 \leq \sigma_s \leq 0.18$~mag. 

\autoref{fig:artstars}b presents the uncertainties associated with edge detection in our datasets as a function of smoothing scales, $\sigma_s$. Squares represent average systematic differences between the measured and input tip magnitude; plusses represent the dispersion of measured tip magnitudes, or a measure of the statistical (random) uncertainty; and circles represent these uncertainties added in quadrature. We find that the location of the TRGB is well-recovered at a $\leq 0.02$~mag level of precision. This result is not entirely surprising: simulations by \cite{2009ApJ...690..389M} have shown that 0.1~mag precision in tip detection can be achieved with $\sim$400 stars in the first magnitude of the RGB. For the galaxies studied here, we have enormous statistical samples of 22224, 15439, and 15069 stars in the first magnitude below the anticipated TRGB. The high-precision in the measurement of the TRGB then suggests that the driving source of uncertainty for the galaxies being studied here is photometric errors. Finally, \autoref{fig:artstars}c shows the distribution of measurements for the simulated TRGBs relative to their input values.

These modeled measurement uncertainties, combined with the adopted calibration errors, show that the galaxies studied here are found to have roughly the same overall uncertainties ($\pm$0.04-0.05~mag) in the measured magnitudes of their tip discontinuities. This fact translates into almost identical uncertainties in their distances ($\pm$0.5~Mpc), scaled to a common TRGB zero-point, as calculated in the following Section.

\subsection{TRGB Measurements and Distances}\label{ssec:dist}

\begin{deluxetable*}{ccccccccccc} 
\tabletypesize{\normalsize} 
\tablewidth{0pt} 
\tablecaption{Summary of TRGB distances to NGC~4424, NGC~4526 and NGC~4536\label{tbl:distances}} 
\tablehead{ 
\colhead{Galaxy} &
\colhead{$m_{\mathrm{TRGB}}$\tablenotemark{a}} &
\colhead{$\sigma_{m}$\tablenotemark{b}} &
\colhead{$A_{\mathrm{F814W}}$} &
\colhead{$m-M$\tablenotemark{c}} &
\colhead{$\sigma_{stat}$} &
\colhead{$\sigma_{sys}$} &
\colhead{$D$~(Mpc)} &
\colhead{$\sigma_{stat}$} &
\colhead{$\sigma_{sys}$} &
}
\startdata 
\ngctwofour & \mTRGBngctwofourROUNDED & \ngctwofourmTRGBerrCMBROUNDED  & \IextinctionngctwofourROUNDED & \ngctwofourDMROUNDED & \ngctwofourCMBstaterrROUNDED & \ngctwofourCMBsyserrROUNDED & \ngctwofourDistROUNDED & \ngctwofourDiststaterrROUNDED & \ngctwofourDistsyserrROUNDED\\
\ngctwosix &  \mTRGBngctwosixROUNDED & \ngctwosixmTRGBerrCMBROUNDED  & \IextinctionngctwosixROUNDED &\ngctwosixDMROUNDED & \ngctwosixCMBstaterrROUNDED & \ngctwosixCMBsyserrROUNDED & \ngctwosixDistROUNDED & \ngctwosixDiststaterrROUNDED & \ngctwosixDistsyserrROUNDED \\
\ngcthreesix &  \mTRGBngcthreesixROUNDED & \ngcthreesixmTRGBerrCMBROUNDED & \IextinctionngcthreesixROUNDED & \ngcthreesixDMROUNDED & \ngcthreesixCMBstaterrROUNDED & \ngcthreesixCMBsyserrROUNDED & \ngcthreesixDistROUNDED & \ngcthreesixDiststaterrROUNDED & \ngcthreesixDistsyserrROUNDED  \\
\enddata 
\tablenotetext{a}{F814W}
\tablenotetext{b}{Combined statistical and systematic uncertainties from the edge detection method and calibration to the \hst flight magnitude system.}
\tablenotetext{c}{$M_I^{\mathrm{TRGB}}=\trgblum\pm0.03\pm0.05$~mag.}
\end{deluxetable*} 

On the topic of line-of-sight reddening, the targets of this study were chosen, in part, because of their estimated low foreground extinction (i.e., $E(B-V)\leq0.02$ \citep[][obtained via the NASA/IPAC Extragalactic Database, NED]{2011ApJ...737..103S}. Adopting a \cite{1989ApJ...345..245C} reddening law, the predicted foreground reddening is $A_{\mathrm{F814W}}=\IextinctionngctwofourROUNDED$~mag in all three cases. The uncertainty in $E(B-V)$ is estimated to be $\pm0.03$~mag \citep{1998ApJ...500..525S}, which suggests that the foreground extinctions for each of these galaxies are statistically consistent with zero. We conservatively include half of the value of the predicted extinction as an additional systematic error in the distance moduli derived below.

Although the foreground reddening is predicted to be very low, it
is not yet possible to assess whether there is extinction intrinsic to the halos themselves. One possible test for the presence of halo dust, however, is to observe whether or not the apparent TRGB magnitude changes with increasing projected distance from the galaxy. We tested for this possibility by dividing the images into two distinct regions of stars having equal numbers. We re-ran our TRGB simulations with the adjusted star counts for this new test to find the appropriate level of GLOESS smoothing to minimize the combined measurement uncertainties. We found that even with the reduced statistics, the required level of smoothing is comparable to its original value with the full stellar catalogs. In each of the three galaxies, we found that the region closest to the disk was in fact \emph{brighter} than further away, ranging between 0.05-0.10~mag. This observation is the opposite expectation to a reddening effect, and we note that the previous two targets in the \cchp series, IC\,1613 and NGC\,1365, had no discernible difference in the measured TRGB across different regions of the imaging. Instead, it suggests that the edge detector may be triggering off bright stars or AGB stars---a known systematic when the number of TRGB stars is limited---or other stellar populations near the TRGB that are not resolved in color-magnitude space due to signal-to-noise considerations and the proximity of the imaging to their respective bodies.

In Section \ref{sec:compl}, we identified blends as probably being present for the $10\%$ of the footprint for \ngctwofour closest to its disk, which is likely to be the source of the brighter TRGB measurement for that galaxy.
We plan to undertake a more comprehensive, consistent review of the photometry for all galaxies that are part of the \cchp in a future work in order to make an informed decision of when regions of a CCD are compromised and, if so, how they can be accounted for. At present, we take the entire photometry sample for \ngctwosix and \ngcthreesix as is for the basis of our science result. As stated in Section \ref{sec:compl}, we exclude the region of \ngctwofour whose photometry is likely compromised by blends.

We turn now to the question of metallicity. At high metallicity, a downward sloping TRGB is observed in color-magnitude space for the reddest stars at optical wavelengths. The observations used in this study were specifically crafted to target the metal-poor halos of these galaxies, however. As a result, the TRGBs in our sample of galaxy halos do not show any discernible color-magnitude dependence and are not red enough to  necessitate the application of TRGB `rectification tools' \citep[for ACS filters see][]{2017ApJ...835...28J}; in other words, there is negligible slope to correct.

\begin{figure*}
\centering
\includegraphics[angle=0,width=0.7\textwidth]{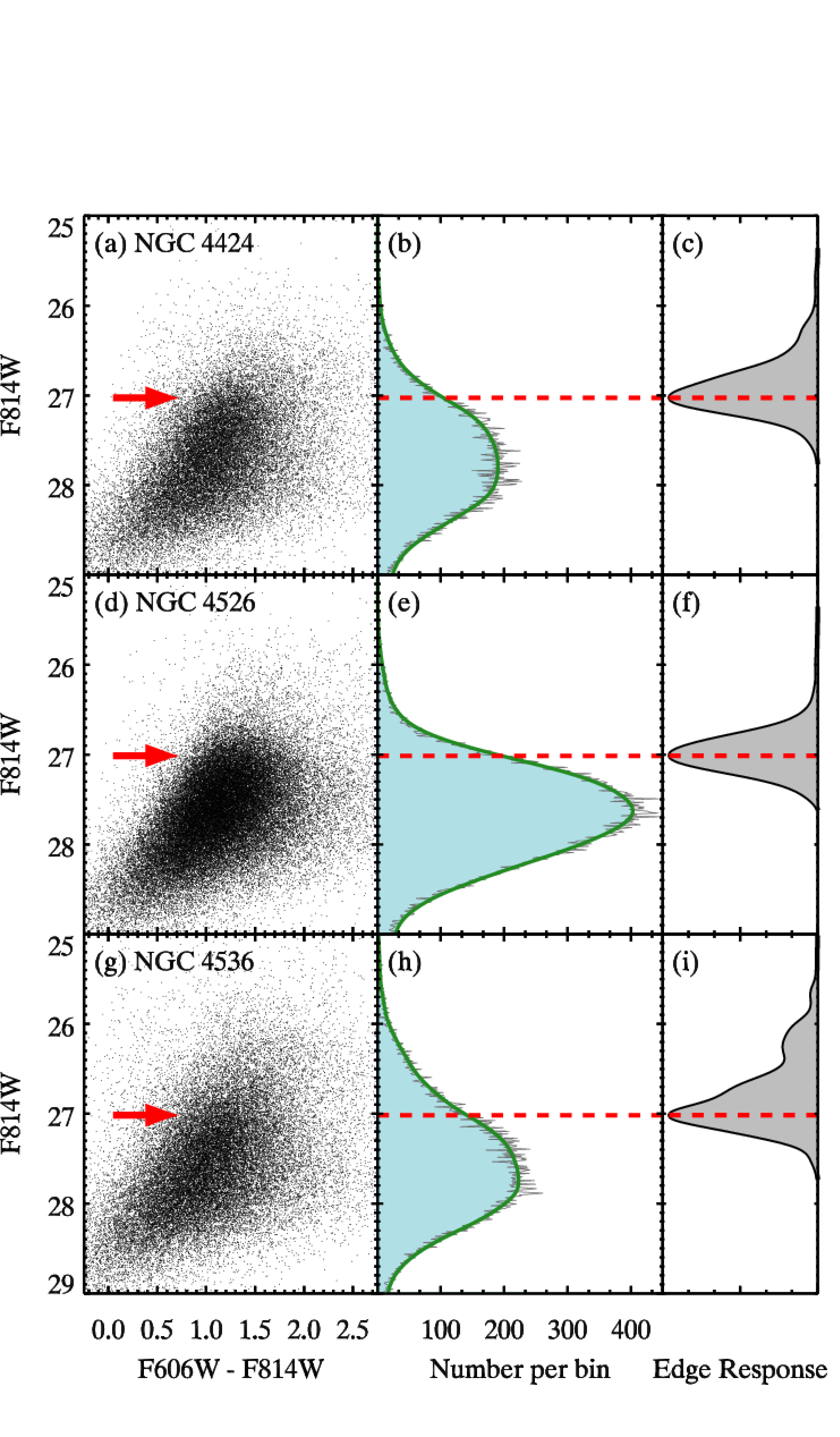}
\vspace{0.0cm}
\caption{TRGB edge detection for \ngctwofour, \ngctwosix, and \ngcthreesix. Panels (a), (d), and (g) display the CMDs. A red arrow in each plot shows the location of the measured TRGB. Panels (b), (e), and (h) show the binned luminosity functions in 0.01 mag intervals in gray and the GLOESS-smoothed luminosity functions in green. Panels (c), (f), and (i) show the response function of the $[-1,0,+1]$ kernel on the smoothed luminosity functions. These functions are scaled in order to align their peaks. A dashed line passes through the location of the greatest change in the luminosity functions.
\label{fig:distances}}
\end{figure*}

\autoref{fig:distances} displays the results of the TRGB measurement using the optimally selected GLOESS-smoothing scales for each galaxy. We find the following F814W ($I$-band equivalent) TRGB magnitudes: For \ngctwofour, $I(TRGB) = $ $\mTRGBngctwofourROUNDED$~mag, for \ngctwosix, $I(TRGB) = $ $\mTRGBngctwosixROUNDED$~mag, and for \ngcthreesix, $I(TRGB) = $ $\mTRGBngcthreesixROUNDED$~mag. The targeting of the galaxy halos minimizes the need for the isolation of RGB stars via a color-magnitude selection cut. We note only a 0.01~mag difference in the aforementioned measurement for \ngctwofour when removing the bluest of sources $\mathrm{F606W}-\mathrm{F814W} \leq 0.6$ within the CMD. As with the study of NGC\,1365 in \citetalias{2018ApJ...852...60J}, all three galaxies have little to no difference in measured TRGB value at the 0.01~mag level with or without a color-magnitude selection cut.

We noted in Section \ref{sec:cmds} that \ngcthreesix has an apparent spiral arm or other stellar feature that passes through the imaging. In order to ensure that we are largely, if not exclusively, measuring the TRGB for Pop II stars, we manually select the physical region in the imaging that encompasses what appears to be the arm. We then compare the difference in the TRGB measurement with and without the contribution of stars in this region. We find that the TRGB is only 0.01~mag fainter without the arm than with it, or in other words, a difference that is negligibly small compared to the other uncertainties we have quoted thus far in the analysis. 

We conclude this section by making explicit our adopted zero-point for the TRGB. The F814W filter is a ``broad $I$'' filter, and the observed transformation from F814W to the $I$-band is a negligible $0.002\pm0.017$~mag effect for a common color of $V-I\approx1.0$ for bright RGB stars \citep[see Table 22 of][]{2005PASP..117.1049S}. We therefore adopt a provisional zero-point in the $I$-band, using three recent determinations of the $I$-band discontinuity of the TRGB in the LMC: (1) \cite{rom00}
give a reddening-corrected value of $I_o$ = 14.50 $\pm$ 0.25 mag.  (2) \cite{sak00} find $I_o$ = 14.54 $\pm$ 0.04 mag. (3) \cite{cio00}
also find $I_o$ = 14.54 $\pm$ 0.04 mag. The weighted mean of $I_{LMC} = 14.54$~mag compares favorably with the redetermination of \cite{2017ApJ...835...28J}, who obtain $I_o$ =
14.524 $\pm$ 0.042 from the mean of eight fields covering the eclipsing binaries with
known geometric distances. We adopt E(B-V) =  0.03$\pm$0.03~mag for the reddening of the LMC TRGB stars \citep[see][]{2018ApJ...858...12H}. Combining
this corrected, weighted mean with our adopted LMC distance modulus of 18.49~mag, we arrive at $M_I =-4.00$~mag $\pm 0.03_{stat}\pm 0.05_{sys}$ for the $I$-band TRGB  absolute zero-point
calibration. At present, we have adopted the systematic uncertainty for the LMC distance following Pietrzynski et al. 2013, based on their study of detached eclipsing binaries in the LMC bar.\footnote{The interested reader is referred to the following four papers for
detailed discussions of the intricacies of calibrating the zero-point of the I-band TRGB,
(a) through the galactic globular cluster $\omega$~Cen \citep{bel01}, (b) using
scaled theoretical models \citep{bel08}, (c) setting the zero-point at the LMC \citep[][who derive a totally consistent zero-point $M_I = -3.970 \pm 0.102$~mag to the
value derived here, albeit with a significantly larger uncertainty]{2017ApJ...835...28J}, and finally (d)
iterating distance determinations to nearby galaxies \citep{riz07}. In the latter
paper, we note that their Figure 13 indicates that the maser NGC~4258 (which has a truly
independent, geometric  distance determination) gives an $I$-band TRGB zero-point of
-3.94~mag. Their preferred value is however, -4.05$\pm$0.02 at $(V-I) = 1.6$~mag.} Applying the corresponding foreground reddening correction and provisional TRGB zero-point, we present the distance moduli and true distances in \autoref{tbl:distances}.

\section{Distance comparisons}\label{sec:dist_compare}

\begin{figure*}
\centering
\includegraphics[angle=0,width=1.0\textwidth]{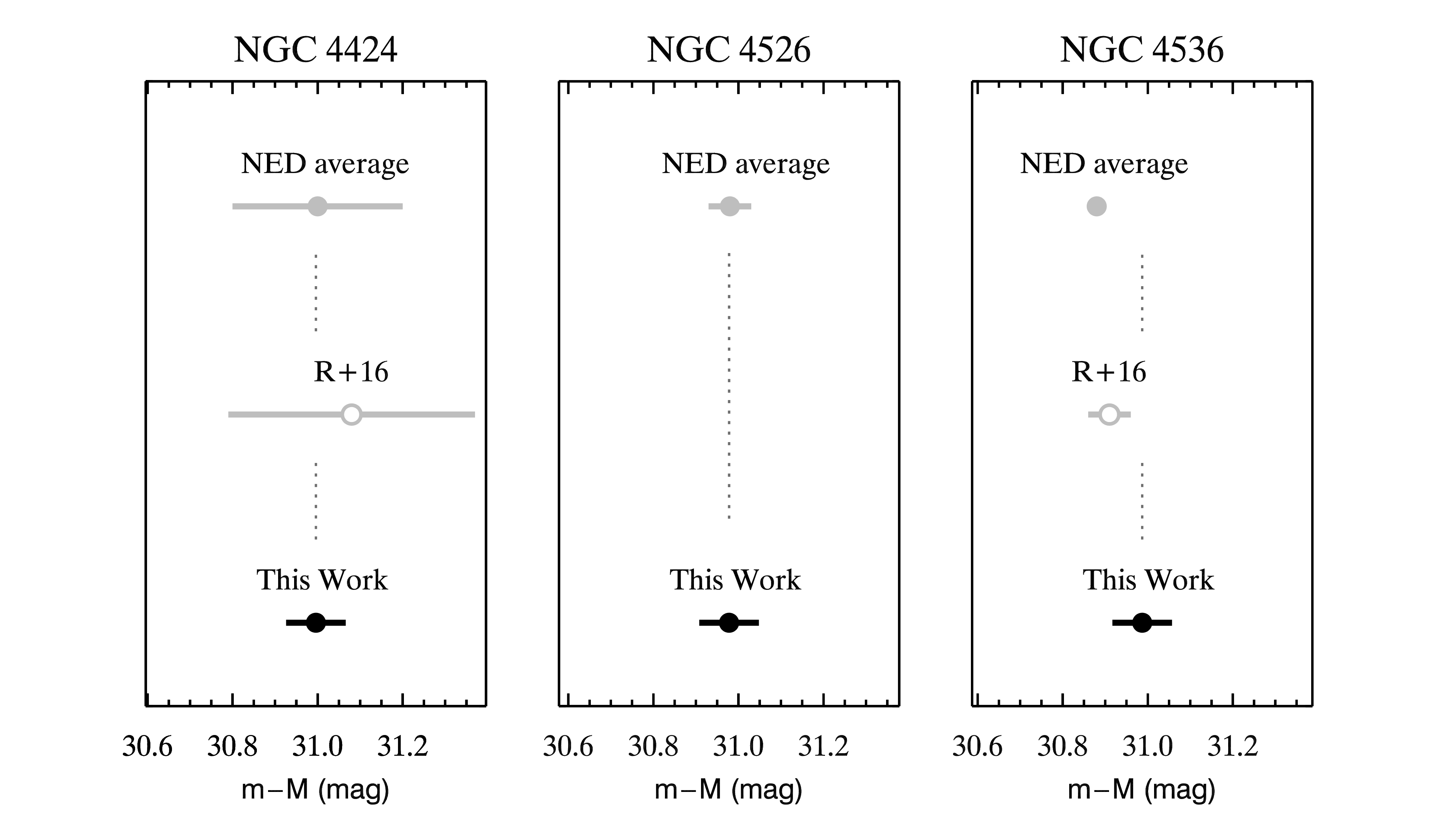}
\vspace{0.0cm}
\caption{Distance moduli and combined statistical (random) and systematic uncertainties as part of this work compared with the most recent Cepheid-based results, where available, and the average of select, recently published distances (accessed through NED). Vertical dotted lines align with the results of this work. Cepheid distances are published as ``approximate'' and are denoted here by open circles \citep[see Table 5 of][labeled here as R+16]{2016ApJ...826...56R}. Error bars on the NED averages are the error on the mean for published distances ``as is'', without adjusting for common zero-points and reddening, which will invariability underestimate the true uncertainty in the average. The TRGB method is able to achieve consistent precision for the comparably distant galaxies presented here, whereas population statistics can dominate the error budget for Cepheids, which is the case for \ngctwofour.
\label{fig:distance_comparisons}}
\end{figure*}

The following are comparisons of the TRGB distances determined here against previously-published, independent measurements, which are visualized in \autoref{fig:distance_comparisons}. 

\subsection{\ngctwofour}

At the time of this writing, there are four publications reporting distances to this galaxy. Two of these estimates are based on the Tully-Fisher relation, though only the more recent analysis by \cite{2008ApJ...683...78C} is based on a direct measurement of \ngctwofour. Their distance of $30.90\pm0.30$~mag, obtained using synthetic $H_\alpha$ rotation curves, is considerably more uncertain compared to our $\ngctwofourDMROUNDED~\pm~\ngctwofourCMBstaterrROUNDED_{stat}~\pm \ngctwofourCMBsyserrROUNDED_{sys} $~mag, but there is mutual agreement in the estimates to within their reported errors.

The third of four publications on the distance to \ngctwofour is based on observations of SN~2012cg itself \citep{2013NewA...20...30M}. They find an average distance of 30.95~mag (no quoted uncertainty and using only $VRI$ observations), which, though consistent with our estimate, is not a truly independent comparison given the SN~Ia calibration objectives of the \cchp.

The fourth and most recent estimate is based on just 3 Cepheids published in \cite{2016ApJ...826...56R}. That paper finds an ``approximate'' distance modulus of $31.08~\pm$ 0.29~mag. To within one sigma in the combined uncertainties, that distance determination for {\ngctwofour} is formally consistent with our more precise value.

The compilation of independent estimates (Tully-Fisher and Cepheids), obtained via NED \citep[excluding][]{2013NewA...20...30M}, yields a mean distance of $31.00~ \pm$ 0.21~mag. This value is only a slight improvement over the two original estimates alone, but it is still consistent with the value determined here.

\subsection{\ngctwosix}

There are over 20 published distance moduli for this galaxy, but the span in those estimates is large, ranging from 29.23 to 31.47~mag. Some of the most precise measurements for \ngctwosix come from Globular Cluster Luminosity Functions (GCLFs) that were derived as part of the the ACS Virgo Cluster Survey \citep{2007ApJS..171..101J,2010ApJ...717..603V}. The most recent estimates presented in \cite{2010ApJ...717..603V} give $30.89\pm0.10$~mag and $31.03\pm0.09$~mag for the $g$- and $z$-bands, respectively, or an average $30.97\pm0.07$~mag. Individually or combined, the GCLFs and our estimate \ngctwosixDMROUNDED ~$\pm \ngctwosixCMBstaterrROUNDED_{stat}\pm \ngctwosixCMBsyserrROUNDED_{sys}$~mag are in mutual agreement.

Additional distances to \ngctwosix are available based on its \sne, SN 1994D. Of the 14 publications that report a \sne distance, 2  report measurements with sub-0.1~mag uncertainties: \cite{2000ApJ...540..634P} and \cite{2007ApJ...659..122J}. The average of these is $31.18\pm0.03$, nearly $0.2$~mag fainter than our result and that of the GCLFs (as well as several standard deviations apart). A direct comparison to \sne distances here is not very informative given the goal of the \cchp to independently calibrate them, but we note the possibility of a large systematic due to assumptions in the value of $H_0$.

Finally, the remaining distance indicators with `modern' (since the year 2000) publications include Tully-Fisher \citep[$31.02\pm0.40$~mag,][]{2012ApJ...749..174C}  and Surface Brightness Fluctuations (SBF) \citep[average $30.99\pm0.07$~mag,][]{2000ApJ...529..745F,2001ApJ...546..681T,2001ApJ...559..584A,2013AJ....146...86T}. Thus, there appears to be excellent consistency between all distance indicators for \ngctwosix (excluding \sne distances).
The mean of the methods listed above gives a precise $30.98\pm0.05$~mag that is consistent with our distance. Importantly, the TRGB distance given here represents an order-of-magnitude improvement in precision over most previously-published, individual estimates.

\subsection{\ngcthreesix}


Aside from its \sne, distance determinations for \ngcthreesix have focused on Cepheids and the Tully-Fisher relation. For Cepheids, the measured distance moduli have varied greatly with bounds of 30.61~mag to 31.24~mag. Those with reported measurement uncertainties produce an average of $30.88\pm0.01$~mag, where the uncertainty (the error on the mean) is minuscule due to the fact that NED lists 45 estimates from 36 unique publications. We note that we have not adjusted these estimates to a common zero-point or reddening and have simply used the values ``as published''. For the purposes of this work, we note them here merely to establish a reference point for the historical efforts to determine the distance to \ngcthreesix. The most recent Cepheid publication reported an ``approximate'' distance of $30.91~ \pm$ 0.05~mag \citep{2016ApJ...826...56R}, which is consistent with the historical average.

These Cepheids-based distances are to be compared to the value determined here using the TRGB, $\ngcthreesixDMROUNDED~\pm~\ngcthreesixCMBstaterrROUNDED_{stat}~\pm \ngcthreesixCMBsyserrROUNDED_{sys}$~mag. Our distance for \ngcthreesix is approximately 1.2 standard deviations from the recent \cite{2016ApJ...826...56R} estimate, and roughly 1.6 standard deviations from the average of historical Cepheid distances. 

In contrast to the Cepheids, the Tully-Fisher relation has produced a notably smaller distance modulus. Its average is $30.83\pm0.04$~mag from 9 publications that report uncertainties (since the year 2000), though it is only approximately one standard deviation from the historical Cepheid average. When combined with the Cepheid distances, the literature again produces an average $30.88\pm0.01$~mag, which is dominated by the large number of Cepheid estimates and their small reported uncertainties. Despite the small uncertainty on the averages, we again caution that the values are simply used ``as is'' without altering their zero-point or reddening assumptions, which may not be possible depending on the level of detail given in each publication. Thus, it would be safe to assume that the uncertainties on the Cepheid and NED averages are underestimates and could be as large as the standard deviations of their values, 0.13~mag and 0.39~mag, respectively. In this scenario, our reported distance for \ngcthreesix is consistent with the literature.

\section{Conclusion}\label{sec:conc}

We have determined the first Tip of the Red Giant Branch distances to three \sne-host galaxies--- \ngctwofour, \ngctwosix, and \ngcthreesix---which are an integral part of an on-going effort by the \cchp to independently set the \sne absolute zero-point 
using (Population~II) TRGB distances.
We find good agreement between these latest results (with independent systematics) in comparison to a number of previously published distances for each of these galaxies. In particular, we find consistency between the distances derived from Pop~I and II stars in \ngctwofour and \ngcthreesix,
where the comparisons were possible.

The TRGB distances determined here are of relatively high-precision at 3\% uncertainty in distance; and in the case of \ngctwofour and \ngctwosix, they represent a substantial improvement over previously-published estimates. For this reason, the results presented here will serve as valuable calibrators for the \sne extragalactic distance scale. With future \emph{Gaia} data releases, the \cchp will be refining the TRGB distance scale locally using Milky Way RGB stars, thereby improving the zero-point accuracy of the TRGB method as a whole and further improving upon the distances estimates reported here. Longer-term, with the pending launch of \emph{JWST}, it will be possible to extend the TRGB method, as demonstrated here using \emph{HST}, to samples of galaxies at even greater distances.


\section*{Acknowledgments}
We thank Peter Stetson for a copy of $\textsc{DAOPHOT}$ as well as his helpful engagement on its usage. 
Support for this work was provided by NASA through Hubble Fellowship grant \#51386.01 awarded to R.L.B. by the Space Telescope Science Institute, which is operated by the Association of  Universities for Research in Astronomy, Inc., for NASA, under contract NAS 5-26555.
Authors MGL and ISJ were supported by the National Research Foundation of Korea (NRF) grant funded by the Korea Government (MSIP) No. 2017R1A2B4004632.
Support for program \#13691 was provided by NASA through a grant from the Space Telescope Science Institute, which is operated by the Association of Universities for Research in Astronomy, Inc., under NASA contract NAS 5-26555.
This research has made use of the NASA/IPAC Extragalactic Database (NED), which is operated by the Jet Propulsion Laboratory, California Institute of Technology, under contract with the National Aeronautics and Space Administration.

\facility{HST (ACS/WFC)}
\software{DAOPHOT \citep{1987PASP...99..191S}, ALLFRAME \citep{1994PASP..106..250S}, TinyTim \citep{2011SPIE.8127E..0JK}}
\vfill\eject

\bibliographystyle{aasjournal}
\bibliography{ms.bib}

\end{document}